\documentclass[prl,preprint,twocolumn,superscriptaddress,floatfix,10pt]{revtex4-1}%
\usepackage{amsfonts,amsmath,amssymb,bm,bbm}
\usepackage{graphicx}
\usepackage{xcolor}
\allowdisplaybreaks
\usepackage[colorlinks=true,urlcolor=blue,anchorcolor=blue,citecolor=blue,filecolor=blue,linkcolor=blue,menucolor=blue,pagecolor=blue,linktocpage=true,pdfproducer=medialab,pdfa=true]{hyperref}
\usepackage{dsfont}
\usepackage{calrsfs}
\usepackage{footmisc}
\usepackage{comment}
\usepackage{xfrac}
\DeclareMathAlphabet{\pazocal}{OMS}{zplm}{m}{n}
\DeclareMathAlphabet{\mathdsl}{U}{bbm}{m}{sl}

\usepackage{braket}

\newcommand{\T}{\mathsf{T}}
\newcommand{\h}{\mathsf{h}}
\newcommand{\e}{\mathsf{e}}
\newcommand{\Q}{\mathsf{Q}}
\newcommand{\sfi}{\mathsf{i}}
\newcommand{\sfj}{\mathsf{j}}

\makeatletter
\DeclareFontFamily{OMX}{MnSymbolE}{}
\DeclareSymbolFont{MnLargeSymbols}{OMX}{MnSymbolE}{m}{n}
\SetSymbolFont{MnLargeSymbols}{bold}{OMX}{MnSymbolE}{b}{n}
\DeclareFontShape{OMX}{MnSymbolE}{m}{n}{
    <-6>  MnSymbolE5
   <6-7>  MnSymbolE6
   <7-8>  MnSymbolE7
   <8-9>  MnSymbolE8
   <9-10> MnSymbolE9
  <10-12> MnSymbolE10
  <12->   MnSymbolE12
}{}
\DeclareFontShape{OMX}{MnSymbolE}{b}{n}{
    <-6>  MnSymbolE-Bold5
   <6-7>  MnSymbolE-Bold6
   <7-8>  MnSymbolE-Bold7
   <8-9>  MnSymbolE-Bold8
   <9-10> MnSymbolE-Bold9
  <10-12> MnSymbolE-Bold10
  <12->   MnSymbolE-Bold12
}{}

\let\llangle\@undefined
\let\rrangle\@undefined
\DeclareMathDelimiter{\llangle}{\mathopen}%
                     {MnLargeSymbols}{'164}{MnLargeSymbols}{'164}
\DeclareMathDelimiter{\rrangle}{\mathclose}%
                     {MnLargeSymbols}{'171}{MnLargeSymbols}{'171}               
\makeatother
\usepackage{stmaryrd}
\DeclareFontFamily{U}{mathx}{\hyphenchar\font45}
\DeclareFontShape{U}{mathx}{m}{n}{<-> mathx10}{}
\DeclareSymbolFont{mathx}{U}{mathx}{m}{n}
\DeclareMathAccent{\widebar}{0}{mathx}{"73}

\DeclareMathAlphabet\mathbfcal{OMS}{cmsy}{b}{n}

\usepackage{etoolbox}
\makeatletter
\patchcmd{\frontmatter@abstractheading}{\abstractname}{}{}{}
\makeatother

\begin{document}

\title{Regularising Spectral Curves for Homogeneous Yang-Baxter strings}
\preprint{\textsc{Prepared for PLB}}

\author{Sibylle Driezen}
\email{sdriezen@phys.ethz.ch}
\affiliation{Institut f\"ur Theoretische Physik, ETH Z\"urich, Wolfgang-Pauli-Strasse 27, 8093 Z\"urich, Switzerland}

\author{Niranjan Kamath}
\email{niranjannkamath@gmail.com}
\affiliation{Institut f\"ur Theoretische Physik, ETH Z\"urich, Wolfgang-Pauli-Strasse 27, 8093 Z\"urich, Switzerland}
\affiliation{Theoretische Natuurkunde, Vrije Universiteit Brussel (VUB) and  The International Solvay Institutes, Pleinlaan 2, B-1050 Brussels, Belgium}

\vspace{20pt}

\begin{abstract}

In   this Letter, we  study the semi-classical spectrum of  integrable worldsheet $\sigma$-models using the Spectral Curve. We consider a Homogeneous Yang-Baxter deformation of the $AdS_5\times S^5$ superstring, understood as the composition of a Jordanian  with a ``non-diagonal'' TsT deformation. We derive its type IIB supergravity solution, whose isometry algebra features zero supercharges and a non-relativistic conformal algebra in $0+1$ dimensions. 
While the Spectral Curves of non-diagonal TsT models are ill-defined, we demonstrate that the composition with a Jordanian model regularises this issue. From the regularised Curve, we  derive the one-loop shift of the classical energy and the semi-classical spectrum of excitations of a point-like string.  In the TsT limit, the one-loop shift vanishes despite the loss of supersymmetry. 
Our results  suggest that it may be possible to use standard Bethe Ansatze on   spin chain pictures of deformed ${\cal N}=4$ Super-Yang-Mills theory dual to non-diagonal TsT models.
\end{abstract}


\maketitle

One   of the earliest and most robust validations of the AdS/CFT correspondence is the match between  the energy spectrum of the $AdS_5\times S^5$  $\sigma$-model and the 
scaling dimensions for various operators of planar ${\cal N}=4$ Super-Yang-Mills (SYM) theory. 
Crucial for this  achievement was the use of an underlying integrable model and  the development of  new integrability techniques to solve for the exact spectrum in certain sectors. 
Particularly well-understood is the spectrum of long operators
dual to semi-classical string configurations
which  are amenable to  the algebraic Bethe Ansatz and the semi-Classical Spectral Curve (s-CSC) respectively, as reviewed in \cite{Beisert:2010jr,*Staudacher:2010jz,*Schafer-Nameki:2010qho}.

The success of AdS/CFT integrability has spurred, among other factors, significant interest in integrable deformations of string $\sigma$-models, cf.~\cite{Hoare:2021dix} for a review.
These deformations preserve classical integrability but break many Noether symmetries, which encourages the development of  new exact techniques  for  non-maximally supersymmetric  generalisations of AdS/CFT. 
While numerous  examples have now been developed, applying  integrable methods to these models has proven challenging.
Notable exceptions are ``diagonal'' T-duality-shift-T-duality (TsT) transformations \cite{Lunin:2005jy,*Frolov:2005dj,*Frolov:2005ty} and inhomogeneous Yang-Baxter deformations \cite{Klimcik:2008eq,*Delduc:2013fga,*Delduc:2013qra}, where techniques similar to those used in the undeformed cases are applicable 
\cite{Alday:2005ww,*Beisert:2005if,*deLeeuw:2012hp,*Kazakov:2018ugh,*Arutyunov:2014wdg,*Klabbers:2017vtw,*vanTongeren:2021jhh,*Seibold:2021rml}.
This can be attributed to the fact that these models preserve the Cartan subalgebra of the original symmetries.

In contrast, generic Homogeneous Yang-Baxter (HYB) deformations \cite{Kawaguchi:2014qwa,*vanTongeren:2015soa}, which include all  TsT-transformations \cite{Osten:2016dvf} as well as non-abelian generalisations, generally break the Cartan subalgebra, rendering existing exact techniques challenging.
Nonetheless, progress has been made using the fact that HYB deformations are on-shell equivalent to the undeformed $\sigma$-model with twisted boundary conditions~\cite{Borsato:2021fuy}.~This allowed the development of the s-CSC for particular point-like string solutions of a non-diagonal TsT model \cite{Ouyang:2017yko} and a non-abelian HYB  deformation of Jordanian type \cite{Borsato:2022drc}. 
However, unlike Jordanian models, the twist for non-diagonal TsT models is non-diagonalisable. This fact 
makes the asymptotics of the curve that holds the energy spectrum of non-diagonal TsT models non-polynomial \cite{Borsato:2021fuy} and thereby turns the  reconstruction of more generic (finite-gap) solutions and their spectra ill-defined.
This limitation is reflected on the field theory side  by rendering usual Bethe Ansatz techniques inapplicable and necessitating more complex methods \cite{Guica:2017mtd} \footnote{Despite this,  more intricate methods based on the Baxter equation  have been successfully employed in \cite{Guica:2017mtd} resulting in the spectrum of a twisted spin chain with which the semi-classical spectrum of a specific point-like string configuration matched at large R-charge \cite{Ouyang:2017yko}.}.

In this Letter, we consider a Jordanian  deformation of  $AdS_5\times S^5$ that combines a \textit{minimal}  Jordanian  with a non-diagonal TsT deformation. 
We show that this composition  regularises the issues related to  the non-diagonalisable twist and non-polynomial asymptotics  of the non-diagonal TsT model. 
We derive the  spectrum of  excitations and the one-loop shift of the classical energy of a  point-like string solution  from this ``regularised'' curve. In the TsT limit, the one-loop shift vanishes, while the degeneracy of  excitations depends on the TsT parameter.

For the corresponding   $\sigma$-models,  we derive the  type IIB supergravity  solution explicitly from the worldsheet and show that the deformed metric is supported by NSNS, $F^{(3)}$ and $F^{(5)}$ fluxes, along with a   constant dilaton.  The minimal Jordanian model  preserves 12 supersymmetries,  the maximum found in the classification of Jordanian deformations of $AdS_5\times S^5$ superstrings \cite{Borsato:2022ubq}, while the  non-diagonal TsT model and their combination do not preserve any supersymmetry. Interestingly, the non-compact sector of the background exhibits non-relativistic conformal isometries encoded by a Schr\"odinger algebra in zero spatial dimension \footnote{Models with Schrödinger symmetries in general dimensions have attracted also much attention as possible holographic descriptions of   anisotropic condensed matter systems \cite{Nishida:2007pj,*Son:2008ye}.}. This is the non-relativistic analogue of the conformal algebra  relevant for the SYK model \cite{Sachdev:1992fk,*KitaevTalk1,*KitaevTalk2} and $AdS_2/CFT_1$ holography \cite{Maldacena:2016hyu,*Sarosi:2017ykf}.
In the non-diagonal TsT limit, the background further simplifies significantly and recovers the isometries of the round $S^{5}$ sphere. 

With the semi-classical spectrum of a point-like string, our work   provides concrete results at the gravity side, which we hope could be matched in the future with a   holographic description of a HYB deformation of ${\cal N}=4$ SYM  or a potential anisotropic  QFT obtained after dimensional reductions. Much progress is being made on the former \cite{vanTongeren:2015uha,*vanTongeren:2016eeb,*Meier:2023kzt,*Meier:2023lku}, where they are understood as noncommutative deformations of SYM defined through twisted field products. However, a concrete construction for the Jordanian cases is  yet to be developed. 

Using the regularisation of a non-diagonal TsT with a Jordanian model, we  anticipate  the  usage of traditional Bethe Ansatz techniques in the spin chain representation of QFTs dual to non-diagonal TsT models, such as e.g.~the dipole deformations \cite{Matsumoto:2015uja,Guica:2017mtd} \footnote{It would be    interesting to  verify compatibility with the results of \cite{Ouyang:2017yko}.}. The non-diagonal TsT regularisation  presented in this Letter can in fact be extended to  other examples in the  classification of Jordanian deformations of $AdS_5\times S^5$ \cite{Borsato:2022ubq} \footnote{One can include Jordanian models that do not admit a unimodular extension, as this feature does not play a role in the non-diagonal TsT limit as explained later.}. It would therefore be valuable to understand and classify the  space of possible regularisations of non-diagonal TsT models within \cite{Borsato:2022ubq}  and to examine,  for example, when there are nontrivial one-loop shifts in the energy more systematically. Another particular interesting example obtained by non-diagonal TsT transformations is the Hashimoto-Itzhaki/Maldacena-Russo background \cite{Hashimoto:1999ut,*Maldacena:1999mh}, understood as the Groenewold-Moyal noncommutative deformation of SYM \cite{Matsumoto:2014gwa,vanTongeren:2015uha,*vanTongeren:2016eeb,*Meier:2023kzt,*Meier:2023lku}.  In this case, taking the non-diagonal TsT limit directly on the twist  will yield a non-diagonalisable result,  thus requiring Jordanian regularisation  at each stage of the computations.

Extending  our s-CSC results to a Quantum Spectral Curve (QSC) description would also be very compelling in order to obtain and match with the exact spectrum of the underlying integrable models of the deformed duals. At the CSC level, 
we find that the deformations only affect the asymptotics of the curve, which aligns with known   QSC descriptions of other  deformations, e.g.~\cite{Gromov:2015dfa,*Kazakov:2015efa,*Gromov:2017cja}. 

Interestingly, the non-diagonal TsT model we consider uses the same deformation operator as in \cite{Idiab:2024bwr}, where it acts on the string $\sigma$-model in flat space instead of  $AdS_5\times S^5$, which allowed the authors to obtain the exact energy spectrum.  Understanding a sensible flat space limit of our model and matching our results with theirs in the semi-classical limit would thus be very interesting \footnote{We thank Stijn Van Tongeren for pointing out this connection.}.

Another intriguing possibility that we wish to understand further is the possible applications of our work to non-relativistic versions of $AdS_2/CFT_1$. 

\vspace{10pt}

\textit{\textbf{The Jordanian string.}}
%
Homogeneous Yang-Baxter deformations of semi-symmetric space $\sigma$-models 
 are realised by
 the action  
\cite{Klimcik:2008eq,*Delduc:2013fga,*Delduc:2013qra}
\begin{equation}%
S = - \frac{\sqrt{\lambda}}{4\pi} \int  d^2\sigma \  \mathrm{str} ( \Pi^{\alpha\beta}  J_\alpha \hat{d}_- ({1-\eta R_g \hat{d}_-})^{-1} J_\beta ) \ , \label{eq:action-hyb}
\end{equation} %
where $\frac{\sqrt{\lambda}}{4\pi}$ denotes the string tension, $\Pi^{\alpha\beta}= \frac{1}{2}(\gamma^{\alpha\beta} - \epsilon^{\alpha\beta})$ with $\gamma^{\alpha\beta}$  the unit worldsheet metric, $\epsilon^{\tau\sigma} = -\epsilon^{\sigma\tau}=-1$, $\mathrm{str}$  the supertrace of a $\mathbb{Z}_4$-graded superalgebra $\mathfrak{g}=\mathrm{Lie}(G)$,  $J=g^{-1}dg$ with $g(\tau, \sigma) \in G$, and $\hat{d}_\pm =\mp \frac{1}{2} P^{(1)} + P^{(2)} \pm \frac{1}{2} P^{(3)}$ with $P^{(0,1,2,3)}$  projectors on the $\mathbb{Z}_4$-eigenspaces. The deformation is induced by $R_g = \mathrm{Ad}_g^{-1} R \mathrm{Ad}_g$, with $\mathrm{Ad}_g x = g xg^{-1}$ for $x\in \mathfrak{g}$ and $R : \mathfrak{g} \rightarrow \mathfrak{g}$ a linear operator, and $\eta \in \mathbb{R}$ is the deformation parameter. When $R$  is antisymmetric with respect to $\mathrm{str}$ and  solves the classical Yang-Baxter equation (CYBE) the $\sigma$-model \eqref{eq:action-hyb} is  integrable  \cite{Klimcik:2008eq,*Delduc:2013fga,*Delduc:2013qra}. When $R$  is also unimodular with respect to $\mathfrak{g}$, the $\sigma$-model \eqref{eq:action-hyb} will give rise to a  type IIB supergravity solution if the $\eta=0$ point does
\cite{Borsato:2016ose}. 

Introducing a basis $\mathsf{T}_{\mathsf{A}}$ for $\mathfrak{g}$, with ${\mathsf{A}}=1, \ldots, \dim \mathfrak{g}$, we can write $R \T_{\mathsf{A}} = R^{\mathsf{B}}{}_{\mathsf{A}} \T_{\mathsf{B}}$ and $K_{\mathsf{AB}} = \mathrm{str}(\T_{\mathsf{A}} \T_{\mathsf{B}})$. The $R$-operator is  often written as a  2-fold (graded) wedge product $r =-\frac{1}{2} R^{\mathsf{AB}} \T_{\mathsf{A}} \wedge \T_{\mathsf{B}}$ where $R^{\mathsf{B}}{}_{\mathsf{A}} = K_{\mathsf{AC}} R^{\mathsf{CB}}$. Unimodular Jordanian $R$-operators of rank-2 are HYB deformations  for which \cite{2004Tolstoy,Borsato:2016ose,vanTongeren:2019dlq,Borsato:2022ubq}%
\begin{equation}%
r = \mathsf{h} \wedge \mathsf{e} - \frac{i}{2} (\mathsf{Q}_1 \wedge \mathsf{Q}_1 + \mathsf{Q}_2 \wedge \mathsf{Q}_2) \ , \label{eq:r-jord-uni}
\end{equation}%
with $\h, \e$ bosonic   and $\Q_1, \Q_2$ fermionic generators satisfying $[\h, \e] = \e$, $[\Q_{\sfi} , \e]=0$, $[\h, \Q_{\sfi} ] = \frac{1}{2} (\Q_{\sfi} - \epsilon_{{\sfi}{\sfj}} a \Q_\sfj)$, $\{\Q_{\sfi}, \Q_{\sfj}\} = -i \delta_{{\sfi}{\sfj}}\e$, 
and $a\in \mathbb{C}$ a free parameter. %
In this paper, we take $\mathfrak{g} = \mathfrak{psu}(2,2|4)$ and the $R$-operator  $\bar{R}_{1'}$ of the classification of \cite{Borsato:2022ubq} on $b=-1/2$, i.e.~
\begin{equation} \label{eq:r-matrix-def}
\begin{alignedat}{3}
\mathsf{h} &= \frac{\mathsf{D}-\mathsf{J}_{03}}{2} + a \mathsf{J}_{12} , \qquad &&~~\mathsf{e} &&= \mathsf{p}_0+\mathsf{p}_3 \ , \\
\mathsf{Q}_1 &= \frac{1}{\sqrt{2}} {\cal Q}_+^{21} , \qquad  &&\mathsf{Q}_2 &&= \frac{i}{\sqrt{2}} {\cal Q}_-^{21} \ .
\end{alignedat}
\end{equation}
with ${\sf D}$ the dilatation operator, ${\sf J}$ the Lorentz and ${\sf p}$ the translation generators of the conformal subalgebra, and ${\cal Q}_\pm$ supercharges. 
For our  conventions and superalgebra realisation, we refer to app.~B and eq.~(2.2) of \cite{Borsato:2022ubq}.

\vspace{10pt}

\textit{\textbf{The non-diagonal TsT limit and regularisation.}}
An interesting limit can be obtained by sending $\eta \rightarrow 0$ 
while keeping $\eta_{\text{\tiny TsT}} \equiv \eta a$ constant. 
We will call this the \textit{non-diagonal TsT limit}. In fact,  the $R$-matrix \eqref{eq:r-jord-uni},\eqref{eq:r-matrix-def} can  be interpreted as the composition of two HYB deformations $R=R_{\text{\tiny J}}+R_{\text{\tiny TsT}}$; 
the ``minimal'' Jordanian deformation $R_{\text{\tiny J}}= R (a=0)$    followed by the abelian deformation $R_{\text{\tiny TsT}}$ along the commuting $\mathsf{J}_{12}$ and $\mathsf{p}_0+\mathsf{p}_3$  residual isometries of  $R_{\text{\tiny J}}$.  
The  latter is equivalent to  the sequence of abelian T-duality--shift--T-duality (TsT) transformations \cite{Osten:2016dvf} along those isometries. 
In the TsT limit,  due to the multiplication of $R$ by $\eta$ in the action,  $R_{\text{\tiny J}}$ is eliminated while $R_{\text{\tiny TsT}}$ survives. 
Interestingly, our $R_{\text{\tiny TsT}}$ is of non-diagonal type: 
at least one of the commuting generators (here $\mathsf{p}_0+\mathsf{p}_3$) is non-diagonalisable. In this case,  the on-shell equivalent twisted model has a non-diagonalisable twist and, therefore, a CSC with non-polynomial asymptotics for which it is unknown whether the full curve can be reconstructed for all finite-gap solutions  \cite{Borsato:2021fuy}.
Generically, this   can in turn limit   the reconstruction of the s-CSC fluctuations of a specified finite-gap solution. 
However,  we will see that this issue can be regularised by considering the Jordanian composition $R=R_{\text{\tiny J}}+R_{\text{\tiny TsT}}$
whose twist is always diagonalisable. One can then take the  non-diagonal TsT limit on the results.

\vspace{10pt}

\textit{\textbf{Type IIB supergravity   and  isometries.}}  
The deformed target space  will be manifestly isometric  under the  subalgebra $\{\T_{\bar{\mathsf{A}}} \} \subset \mathfrak{psu}(2,2|4)$ satisfying $\mathrm{ad}_{\T_{\bar{\mathsf{A}}}} R = R \mathrm{ad}_{\T_{\bar{\mathsf{A}}}} $ \footnote{Following \cite{Idiab:2024bwr}, we refer to \textit{manifest} isometries as those corresponding to Noether symmetries that leave the $R$ operator invariant and therefore  the Lagrangian manifestly invariant, without compensating total derivatives. 
There may, however, be more general ``enhanced'' symmetries that leave the Lagrangian invariant up to total derivatives.
For realisations of the latter see \cite{Idiab:2024bwr}.}.
This  can be  divided into sets of bosonic generators of the conformal algebra $\mathfrak{t}_{\mathfrak{a}} \subset \mathfrak{so}(2,4)$ and  ${\sf R}$-symmetry algebra $\mathfrak{t}_{\mathfrak{s}} \subset \mathfrak{so}(6)$, and of fermionic supercharges $\mathfrak{t}_{\mathfrak{q}}$. For the unimodular $R$-operator \eqref{eq:r-matrix-def} with generic $\eta, a$ we have  \cite{Borsato:2022ubq}
\begin{alignat}{4}
 \mathfrak{t}_{\mathfrak{a}}  &= \text{span}( \mathsf{D}+\mathsf{J}_{03}, \mathsf{k}_0+\mathsf{k}_3, \mathsf{p}_0, \mathsf{p}_3, \mathsf{J}_{12})  \cong \mathfrak{sl}(2,R) \oplus \mathfrak{u}(1)^2 
  , \nonumber \\
 \mathfrak{t}_{\mathfrak{s}} &= \text{span} ( {\sf R}_{16} - {\sf R}_{24}, \ {\sf R}_{14} + {\sf R}_{26}, {\sf R}_{36} +{\sf R}_{45},  \nonumber \\
 & \qquad\quad\ \ {\sf R}_{34}+{\sf R}_{56}, {\sf R}_{13}+{\sf R}_{25}, {\sf R}_{15}-{\sf R}_{23}, \label{eq:res-iso}\\ &  \qquad\quad\ \  {\sf R}_{12}, {\sf R}_{35}, {\sf R}_{46}) \cong \mathfrak{su}(3) \oplus \mathfrak{u}(1)    \ , \nonumber 
\end{alignat}
and no supercharges $\mathfrak{t}_{\mathfrak{q}} = \emptyset$. 
On the special point $a=0$,  studied in \cite{vanTongeren:2019dlq,Borsato:2022drc},   $\mathfrak{t}_{\mathfrak{q}}$ enhances to 12 supercharges, while $\mathfrak{t}_{\mathfrak{a}}$ and $ \mathfrak{t}_{\mathfrak{s}}$ are as in \eqref{eq:res-iso}.
 In the non-diagonal TsT limit   $\mathfrak{t}_{\mathfrak{s}}$ enhances to $\mathfrak{so}(6)$, while $\mathfrak{t}_{\mathfrak{q}} = \emptyset$ and $\mathfrak{t}_{\mathfrak{a}}$ is as in \eqref{eq:res-iso}.  In the undeformed limit  one of course restores the full $\mathfrak{psu}(2,2|4)$ with the maximal 32 supercharges. 

The isometry algebra $\mathfrak{t}_{\mathfrak{a}} $ corresponds to the Schr\"odinger algebra in zero spatial dimensions, which is the non-relativistic analogue of zero-dimensional conformal symmetry, extended with the central element ${\sf J}_{12}$, which is a remnant of $\mathfrak{so}(4,2)$.
The  $\mathfrak{sl}(2,R)=\text{span}(\mathsf{D}+\mathsf{J}_{03}, \mathsf{k}_0+\mathsf{k}_3, \mathsf{p}_0- \mathsf{p}_3)$ subalgebra of the  Schr\"odinger algebra (in any spatial dimension) is central in defining non-relativistic scaling dimensions, primary operators and a state-operator map and, consequently,  non-relativistic holography \cite{Nishida:2007pj,*Son:2008ye}, while the central element $\mathsf{p}_0+ \mathsf{p}_3$ has the interpretation of non-relativistic mass. 

Let us now extract the IIB supergravity background of the $\sigma$-model \eqref{eq:action-hyb} with the unimodular Jordanian $r$-matrix \eqref{eq:r-jord-uni} and \eqref{eq:r-matrix-def} by following the methods of \cite{Borsato:2016ose}. For this purpose, we can take a  bosonic coset representative parametrised as $g = g_{\mathfrak{a}} g_{\mathfrak{s}}$ with $g_{\mathfrak{a}} \in SO(2,4) $  and $g_{\mathfrak{s}} \in SO(6) $. As the bosonic part of the $R$-operator only affects the $AdS$ space, we will  take ${g}_{\mathfrak{a}}$ in such a way that the three Cartan generators of the residual $\mathfrak{t}_{\mathfrak{a}}$ isometries are realised as shifts through global left-acting transformations $g \rightarrow g_L g$ with $g_L \in G$ constant. We take
\begin{equation}
g_{\mathfrak{a}} =  e^{ T {\sf H}_T + V {\sf H}_V + \Theta {\sf H}_\Theta }  e^{ P {\sf p}_1} e^{\log(Z) {\sf D}} ,
\end{equation}
with ${\sf H}_T, {\sf H}_V$ and ${\sf H}_\Theta$  the Cartan generators given by ${\sf H}_T = \frac{1}{\sqrt{2}} (\mathsf{p}_0 - \mathsf{p}_3 - \frac{1}{2}(\mathsf{k}_0 + \mathsf{k}_3))$, ${\sf H}_V = \frac{1}{\sqrt{2}}(\mathsf{p}_0 + \mathsf{p}_3)$ and ${\sf H}_\Theta = \mathsf{J}_{12}$.  ${\sf H}_T$ is, up to conjugation, the unique time-like Cartan generator of $\mathfrak{t}_{\mathfrak{a}}$   \cite{Borsato:2022drc}. The background is then invariant under shifts of the coordinates $T, V,$ and $\Theta$. $P$ and $Z$ are the remaining ${SO(2,4)}/{SO(1,4)}$ coordinates.  We parametrise $g_{\mathfrak{s}}$ as in app.~C of \cite{Borsato:2015ghf}, with the  ${SO(6)}/{SO(5)}$  coordinates  labelled as $(\phi_1, \phi_2, \phi_3, \xi \equiv \arcsin \omega, r)$. We will denote the collection of all coordinates by $X$.

Following \cite{Borsato:2016ose} (see also \cite{Hoare:2018ngg}),  we then take the operators $O_\pm = 1\pm \eta R_g \hat{d}_{\pm}$ and calculate the one-forms $A_\pm = O_\pm^{-1} J$. 
Observing that for $M=O_-^{-1} O_+$ one has $M^T P^{(2)} M = P^{(2)}$ \footnote{Note that $\hat{d}_{+}=(\hat{d}_-)^T$.}, 
shows that $P^{(2)}MP^{(2)}$ implements a local Lorentz transformation on the grade-2 subspace of $\mathfrak{g}$. This can be realised as $P^{(2)}MP^{(2)} = \mathrm{Ad}_{h}^{-1} P^{(2)} = P^{(2)} \mathrm{Ad}_h^{-1}$ for an element $h\in G^{(0)} = \exp(P^{(0)} \mathfrak{g})$. Hence, we can write
\begin{equation} \label{eq:h-A}
(P^{(2)} A_+) h = h (P^{(2)}A_-)  \ .
\end{equation}
Next, we introduce the bosonic vielbeins of the deformed and undeformed models as $E = P^{(2)} A_+$ and $e=P^{(2)} J$ respectively. Defining the subsets $\T_{\sf a} = P^{(2)} \T_{\sf A} $, $\T_{\alpha_1} = P^{(1)} \T_{\sf A} $, and $\T_{\alpha_2} = P^{(3)} \T_{\sf A} $, with ${\sf a}=0,\ldots, 9$, and $\alpha_{1,2}=1, \ldots , 16$,  we can then write the metric, B-field, and dilaton as $ds^2 = E^{\sf a} E^{\sf b} K_{{\sf a}{\sf b}}$, $B=\frac{1}{2} (O_-^{-1})_{{\sf a}{\sf b}} e^{\sf a} \wedge e^{\sf b}$ and $e^\phi = (\det O_+)^{-1/2}$ respectively, while the  RR-fluxes can be obtained from projecting the RR-bispinor ${\cal S}^{\alpha_1 \beta_2}=8i (\mathrm{Ad}_h (3-4 O_+^{-1} ))^{\alpha_1}{}_{\gamma_1} K^{\gamma_1 \beta_2}$ on the relevant basis elements of the ten-dimensional Clifford algebra; see \cite{Borsato:2016ose} for more details.  For the latter calculation, one can compute $\mathrm{Ad}_h$ by means of the formula (6.8) of \cite{Borsato:2016ose}. This is, however, a  heavy evaluation; we found it more efficient to construct a generic matrix   $h$ and solve for its elements by pulling the linear equation \eqref{eq:h-A} onto the target-space basis one-forms. Demanding that  the result is an element of $G^{(0)}$ then exhibits a unique expression for $h$. We  find
\begin{align}
ds^2 ={}& - \frac{2Z^4(Z^2+P^2)+\eta^2 (Z^2 + (1+4 a^2)P^2) }{2Z^6}dT^2 \nonumber\\& + \frac{dZ^2 + dP^2 + P^2 d\Theta^2 - 2dT dV}{Z^2} + ds^2_{S^5} \ , \nonumber \\
B ={}& \frac{\eta}{\sqrt{2}} \left(\frac{P dP \wedge dT + 2 a P^2 d\Theta \wedge dT + Z dZ \wedge dT}{Z^4} \right)  ,\nonumber \\
F^{(3)} ={}& F^{(3)}_{a=0} + e^{- \phi_0} 4 \sqrt{2}\eta a  \frac{ P dP \wedge dT \wedge dZ  }{Z^5}\,\label{eq:IIB-bgfields} \\
=&{} e^{- \phi_0} \frac{\sqrt{2} \eta }{Z^5} d T \wedge \left[2 P^2 d\Theta \wedge dZ + Z P dP \wedge d\Theta  \right.  \nonumber\\ &  - 4 a P dP \wedge dZ + Z^2 d\phi_3 \wedge ((1-r^2)dZ + rZ dr  ) \nonumber\\ & \left.  + r \omega Z^2 d\phi_2 \wedge  ( \omega Z dr -r \omega dZ + r Z d\omega)   \right. \nonumber
\\ & \left. + rZ^2 (1-\omega^2) d\phi_1 \wedge ( Z dr - r dZ)    \right. \nonumber \\ & \left. - r^2 Z^3 \omega d\phi_1 \wedge d\omega \right] , \nonumber\\
F^{(5)} ={}& F^{(5)}_{\eta =  0} = e^{- \phi_0}(1+\star)  \frac{4P dT \wedge dV \wedge dZ \wedge d P \wedge d\Theta}{Z^5} \ , \nonumber
\end{align}
 $F^{(1)} = 0$, and the dilaton   $\phi = \phi_0$  constant \footnote{The last term in the $B$-field may be removed by the gauge transformation $B \rightarrow B + \frac{\eta}{2 \sqrt{2}}d(Z^{-2} dT) $}. 
 In the minimal Jordanian limit $a\rightarrow 0$, the background must coincide with eq.~(44) of \cite{vanTongeren:2019dlq} up to redefining  $\eta$ and  after performing the (inverse) coordinate transformation (2.20) of \cite{Borsato:2022drc}  \footnote{We find, however, a difference in the expressions of $F^{(3)}_{z \phi_2 x^-}$ and $F^{(3)}_{z \theta x^-}$ which turn out to be typographical errors in eq.~(44) of \cite{vanTongeren:2019dlq}, and we thank Stijn Van Tongeren for confirming this.  We have checked  that \eqref{eq:IIB-bgfields} solves the type IIB field equations and Bianchi identities.}.  In the non-diagonal TsT limit, the $F^{(3)}$ flux  remains non-vanishing but simplifies significantly (with only legs in $AdS$). On $\eta\rightarrow 0$ we naturally find the undeformed $AdS_5\times S^5$ spacetime with $F^{(3)}=0$. 

It is known that the deformed $AdS_5$ metric of \eqref{eq:IIB-bgfields} is geodesically complete on the (formal) parameter surface  $(1+4 a^2) = 0$ (corresponding to the Schr\"odinger spacetime $Sch_2$) \cite{Blau:2009gd} and on $a=0$ \cite{Borsato:2022drc}. We  have checked that this remains to be the case for generic $(\eta, a)$. In fact, only the geodesic equation for the isometric coordinate $V$ is modified, which does not affect the behaviour around the potential pathological  points $Z,P = \{0,\infty\}$. The isometric coordinate $T$ is thus a \textit{global} time-like coordinate.

Let us now consider a  point-like string solution of the $\sigma$-model in the  target space \eqref{eq:IIB-bgfields}, which is the analog of the BMN solution in undeformed $AdS_5 \times S^5$, on which we will apply the s-CSC techniques. We can  take  the ``BMN-like'' solution of \cite{Guica:2017mtd,Borsato:2022drc} which is trivial in the  $P$-direction, and therefore  also valid on non-trivial $a$;%
 \begin{equation}%
 T = a_T \tau , ~~~V = - \frac{\eta^2 a_T}{2 b_Z^2}\tau , ~~~ Z = b_Z , ~~~ \phi_3 = a_\phi \tau, \label{eq:bmn-like-sol}
 \end{equation}%
with $a_T, b_Z, a_\phi$ real constants, $\gamma^{\alpha\beta}=\mathrm{diag}(-1,+1)$,  and  all other fields (bosonic and fermionic) vanishing. The Virasoro constraints are solved on  
$ a_\phi = \sqrt{1-\frac{\eta^2}{2 b^4_Z}} a_T$. 
We will from now on set $b_Z^4=1/2$  so that they require $-1 < \eta<1$.
The Noether Cartan charges  $Q_\bullet = \frac{\sqrt{\lambda}}{2\pi}\int^{2\pi}_0 d\sigma\ \mathrm{str} ( {\sf H}_{\bullet}  \cdot  \mathrm{Ad}_g A^{(2)}_\tau )  $
associated to the $\mathfrak{t}_{\mathfrak{a}}$ symmetries of the  BMN-like solution \eqref{eq:bmn-like-sol} evaluate to \footnote{The full family of Noether (super)charges is given by 
$
  Q_{T_{\bar{\sf A}}} = \frac{\sqrt{\lambda}}{2\pi}\int^{2\pi}_0 d\sigma\ \mathrm{str} \left[T_{\bar{\sf A}}  \cdot  \mathrm{Ad}_g (A^{(2)}_\tau - \frac{1}{2} (A^{(1)}_\sigma-A^{(3)}_\sigma))  \right]
 $
  with $T_{\bar{\sf A}} \in \{ \mathfrak{t}_{\mathfrak{a}}, \mathfrak{t}_{\mathfrak{s}}, \mathfrak{t}_{\mathfrak{q}} \}$.
  }
\begin{equation}
  Q_T = -\sqrt{\lambda}a_T , \quad Q_V = -\sqrt{2\lambda}a_T  , \quad Q_\Theta = 0 \ .
\end{equation}

\vspace{10pt}

\textit{\textbf{Twisted formulation.}}
In the previous sections, we described the  deformed \textit{periodic}  $\sigma$-model. We will now employ the fact that, on-shell, HYB models are classically equivalent  to the undeformed model  with twisted boundary conditions   \cite{Borsato:2021fuy}. For a review  for \textit{Jordanian} HYB models we refer to sec.~4 and 7 of \cite{Borsato:2022drc}. In the following, we will use tildes to denote objects related to the twisted variables.
Note that, for the $R$-operator \eqref{eq:r-jord-uni}--\eqref{eq:r-matrix-def}, the map between  the deformed periodic variables $g(X)$ to the undeformed twisted variables $\tilde g (\tilde{X})$ is only non-trivial in the AdS-sector.  On the  solution \eqref{eq:bmn-like-sol} it results in 
\begin{equation} \label{eq:bmn-like-sol-twisted}
  \tilde{T} = a_T \tau, ~~~\tilde{Z}=\exp \left(\eta a_T \sigma \right) b_Z, ~~~{\phi}_3 =  a_\phi \tau  \ ,
\end{equation}
and all other fields vanishing. This solution satisfies the following  twisted boundary conditions \footnote{In terms of the group element, the twisted boundary conditions read as $\tilde{g}(\tau, 2\pi) = W \tilde{g}(\tau, 0)h$ with $h\in G^{(0)}$ a possible right-acting gauge ambiguity, for more details see \cite{Borsato:2022drc}.}
\begin{equation}
  \tilde{G}(\tau, 2\pi ) = W \tilde{G} (\tau, 0) W^t , ~~~ W = \exp \left(4\pi \eta a_T \h \right) \ ,
\end{equation}
which are written in terms of the gauge-invariant collection of fields $\tilde{G} = \tilde g K  \tilde{g}^t$ with $K$ the $SO(1,4)\times SO(5)$ invariant. On $\eta=0$, $W=1$ and the boundary conditions become periodic.  For generic $\sigma$-model solutions, the twist of the bosonic undeformed fields reads
\begin{equation} \label{eq:twistgen}
  W = \exp \left( \mathbf{Q} (\h - \mathbf{q}\e) \right) \ ,
\end{equation}
with the expressions for $\mathbf{Q}$ and $\mathbf{q}$ in terms of $\tilde g (\tilde X)$ given in eqs.~(4.5) and (4.6) of \cite{Borsato:2022drc}. Our (twisted) BMN-like solution is thus characterised by  $\mathbf{Q} = 4\pi \eta a_T$ and $\mathbf{q}=0$. 

Since on-shell $\mathbf{Q}$ and $\mathbf{q}$ are constant, and in particular time-independent, these objects  correspond to conserved quantities of the twisted model \cite{Borsato:2021fuy}. Yet, their existence  is not apparent from a continuous symmetry of the action, in the traditional sense of Noether's theorem. Importantly, however, for solutions with $\mathbf{Q} \neq 0$, the object $\mathbf{q}$  can  be removed via a suitable field redefinition of $\tilde g$ \cite{Borsato:2022drc}, while  $\mathbf{Q}$ remains a physical ``charge''; In fact, we will see  that $\mathbf{Q}$ characterises the spectrum of the twisted model. We will henceforth refer to $\mathbf{Q}$ as  the ``twist charge''.

The Noether symmetries of the twisted model, on the other hand,  are  generated in the AdS sector by%
\begin{equation}%
  \tilde{\mathfrak{t}}_{\mathfrak{a}}= \{ {\sf D}+ {\sf J}_{03}  , {\sf k}_0 + {\sf k}_3, {\sf p}_0 - {\sf p}_3, {\sf J}_{12} \} \cong \mathfrak{sl}(2,R) \oplus \mathfrak{u}(1) . 
\end{equation}%
This is generally a subset of the symmetry algebra $\mathfrak{t}_{\mathfrak{a}}$ of the deformed model \cite{Borsato:2022drc}. In particular, the rank is reduced: the Cartan of $\tilde{\mathfrak{t}}_{\mathfrak{a}}$  is spanned by $\{{\sf H}_T,{\sf H}_{\Theta}\}$, while  the Cartan of ${\mathfrak{t}}_{\mathfrak{a}}$ is spanned by $\{{\sf H}_T, {\sf H}_\Theta,{\sf H}_V\}$. Morally, as made more precise later,  the third  Cartan charge ${Q}_V$ of the deformed model is thus replaced by the twist charge $\mathbf{Q}$ of the twisted model. The remaining Cartan  charges 
$\tilde{Q}_{\bullet} = \frac{\sqrt{\lambda}}{2\pi}\int^{2\pi}_0 d\sigma \ \mathrm{str}  ( {\sf H}_{\bullet}  \cdot  \mathrm{Ad}_{\tilde{g}} \tilde{J}^{(2)}_\tau  )$ 
coincide on-shell with those of the deformed model \footnote{The full family of Noether charges of the twisted model is given by 
$
  {\tilde Q}_{T_{\tilde{\sf A}}} = \frac{\sqrt{\lambda}}{2\pi}\int^{2\pi}_0 d\sigma\ \mathrm{str} \left[T_{\tilde{\sf A}}  \cdot  \mathrm{Ad}_{\tilde{g}} (\tilde{J}^{(2)}_\tau - \frac{1}{2} (\tilde{J}^{(1)}_\sigma-\tilde{J}^{(3)}_\sigma))  \right]
 $
  with $T_{\tilde{\sf A}} \in \mathfrak{psu}(2,2|4)$ satisfying $W T_{\tilde{\sf A}} W^{-1}=T_{\tilde{\sf A}} $ \cite{Borsato:2021fuy,Borsato:2022drc}.
  }; i.e.~$\tilde{Q}_T = Q_T = -\sqrt{\lambda}a_T$ and $\tilde{Q}_\Theta = Q_\Theta = 0$. 
 
While the solution and its  charges are independent of $a$, the generic twist $W$ \eqref{eq:twistgen} is $a$-dependent through  $\h$. 
Interestingly, in the non-diagonal TsT limit,  $W$ remains diagonalisable and does not coincide with the non-diagonalisable twist $W_{\text{\tiny TsT}}$ of \cite{Borsato:2021fuy} that one would obtain if calculated directly for the non-diagonal TsT model, i.e.~$W_{\text{\tiny TsT}}= \exp (\tfrac{\eta_{\text{\tiny TsT}}}{\sqrt{\lambda}}(  \tilde{Q}_{{\sf J}_{12}} \e -\tilde{Q}_{\e} {\sf J}_{12} ))$.
Instead $W_{\text{\tiny TsT, reg}}\equiv\lim_{\eta\rightarrow 0,  a \rightarrow \eta_{\text{\tiny TsT}}/\eta} W =  \exp(-\frac{\eta_{\text{\tiny TsT}}}{\sqrt{\lambda}}\tilde{Q}_{\e} {\sf J}_{12})$, which is indeed diagonalisable \footnote{This is, however, a special feature of the model under consideration, particularly when the non-diagonal TsT model has \textit{at most} one  non-diagonalisable generator in the $R$-operator. For non-diagonal TsT models with both non-diagonalisable generators in the $R$-operator, such as the one corresponding to the Maldacena-Russo background \cite{Matsumoto:2014gwa}, the non-diagonal TsT limit  would still result in a non-diagonalisable twist. In general we believe that the TsT limit should be taken only on physical quantities such as the energy spectrum.}. 
 Note, furthermore, that for TsT (i.e.~abelian HYB) models $Q_{T_{\bar{\sf A}}}= \tilde{Q}_{T_{\tilde{\sf A}}}$ \cite{Borsato:2021fuy}  and since ${\sf e}$ is proportional to ${\sf H}_V$, the Noether charge $Q_V$ reinstates itself explicitly through the twist charge $\mathbf{Q}_{\text{\tiny TsT, reg}}$. 

\vspace{10pt}

\textit{\textbf{Semi-Classical  Spectral Curve.}}
Both the deformed and twisted $\sigma$-model are classically integrable; each have a flat Lax connection that coincides on the on-shell map between the models. We can therefore employ the methods of the Classical Spectral Curve (CSC) to analyse the spectrum of infinite charges. In contrast to the deformed  model, in the twisted variables the CSC can be fully reconstructed in terms of \textit{local} conserved charges which include the target-space energy $E\equiv\tilde{Q}_T$. To obtain the semi-classical quantum fluctuations and the one-loop shift to the energy of the BMN-like vacuum we will thus    work in the twisted variables.   

The twisted CSC is obtained from the conserved eigenvalues $\lambda(z)$ of the twisted monodromy matrix \cite{Borsato:2021fuy}
\begin{equation}
  \Omega_W (z) = W^{-1} {\cal P}\exp \left(-\int^{2\pi}_0 d\sigma {\cal L}^{\tilde{g}}_\sigma (z) \right) ,
\end{equation}
with ${\cal L}^{\tilde g}(z) = \tilde g {\cal L}(z) \tilde{g}^{-1} -d \tilde g \tilde g^{-1} $  the  gauge-transformed Lax connection of the undeformed supercoset model 
\begin{equation}
  {\cal L} (z) = \tilde{J}^{(0)} +(z_+  + z_- \star)  \tilde{J}^{(2)}  + z \tilde{J}^{(1)} + z^{-1} \tilde{J}^{(3)} \ ,
\end{equation}
with $\tilde{J}^{(i)}\equiv P^{(i)}\tilde{J}$   and $z_\pm \equiv \frac{1}{2}(z^2 \pm z^{-2})$ \footnote{The Hodge star on one-forms is defined on the worldsheet $\Sigma$ as $\star J = J_\alpha \epsilon^{\alpha\beta} \gamma_{\beta\gamma}d\sigma^\gamma$.}.
Since ${\cal L} (z)$ is    flat $ \forall z\in \mathbb{C}$ and periodic also in the twisted variables \cite{Borsato:2021fuy,Borsato:2022drc}, the eigenvalues of $\Omega_W(z)$ give rise to infinite towers of  conserved charges by expanding around fixed values of $z$. Generically, these charges are non-local, but since ${\cal L}^{\tilde g}_\sigma (z=1) =0$, the leading asymptotics  around  $z=1$ are local charges. Redefining the spectral parameter as $z=\sqrt{\frac{x+1}{x-1}}$ \footnote{Note that there is a typographical error in \cite{Borsato:2022drc} in the redefinition from $z$ to $x$ after eq.~(5.3), where $z=\sqrt{\frac{1+x}{1-x}}$ was written instead of $z=\sqrt{\frac{x+1}{x-1}}$.} this point in $\mathbb{C}$ corresponds to $x=\infty$ around which the monodromy matrix reads
\begin{equation}
  \Omega_W (x) = W^{-1} ( 1 + 4\pi \ x^{-1} \ O_W) + {\cal O}(x^{-2}) ,
\end{equation}
with $O_W = \frac{1}{2\pi}\int^{2\pi}_0 d\sigma \   \mathrm{Ad}_{\tilde{g}} (\tilde{J}^{(2)}_\tau - \frac{1}{2} (\tilde{J}^{(1)}_\sigma-\tilde{J}^{(3)}_\sigma)) $. Note that in the AdS-sector $O_W$ is only conserved in the projections on  $\tilde{\mathfrak{t}}_{\mathfrak{a}}$ \footnote{This is in contrast to the undeformed periodic model ($W=1$) where all  projections on the $\mathfrak{psu}(2,2|4)$ generators are conserved.}. However, after diagonalisation and a possible conjugation to the Cartan, we will  precisely uncover the conserved charges only. In fact, in terms of  the AdS \textit{quasimomenta} ${p}_{\hat{1}-\hat{4}}(x)$ obtained from the AdS eigenvalues $\lambda_{\hat i} (x)= e^{i{p}_{\hat{i}}(x)}$ of $\Omega_W(x)$, we obtain \footnote{Note that the quasimomenta have a $2\pi\mathbb{Z}$ ambiguity.}
\begin{equation} \label{eq:QM-offshell} 
  \begin{pmatrix}
     p_{\hat 1} \\
     p_{\hat 2} \\
     p_{\hat 3} \\
     p_{\hat 4} 
  \end{pmatrix}
  \sim{} {\scalebox{1.2}{$\frac{\mathbf{Q}}{2}$}} \begin{pmatrix}
    a \\ 
    i-a \\
    -i - a \\
    a
  \end{pmatrix} 
  + {\scalebox{1.2}{$\frac{2\pi}{x {\sqrt{\lambda}}}$}}
  \begin{pmatrix}
    \tilde{Q}_\Theta - {E} \\  (2i a -1)\tilde{Q}_\Theta \\  (2i a+1)\tilde{Q}_\Theta \\\tilde{Q}_\Theta + {E}
  \end{pmatrix}  .
\end{equation}
It is interesting to compare this situation to the undeformed periodic model ($W=1$), where these asymptotics are dictated by  the three Noether Cartan charges of the $AdS_5$ isometries.  In contrast, in the twisted case the Noether Cartan algebra is only two-dimensional, but the asymptotics of ${p}_{\hat i}(x)$  are still determined by three  conserved quantities: the  Noether  charges $\tilde{Q}_\Theta$ and $E= \tilde Q_T$ and, crucially,  the twist charge $\mathbf{Q}$, similar as in  \cite{Borsato:2021fuy,Borsato:2022drc}.

The quasimomenta of  the BMN-like solution \eqref{eq:bmn-like-sol}, or equivalently \eqref{eq:bmn-like-sol-twisted},  can be evaluated on the full complex plane. In the AdS sector, we obtain
\begin{equation} \label{eq:ads-qm-bmn}
  \begin{aligned}
    \frac{{p}_{\hat 1} (x) }{2\pi  a_T} - a \eta&= - \frac{{p}_{\hat 4} (x) }{2\pi  a_T} + a\eta   = \frac{ \sqrt{x^2-\eta^2}}{x^2-1} ,\\
    \frac{{p}_{\hat 2} (x) }{2\pi  a_T} + a\eta &= - \frac{{p}_{\hat 3} (x) }{2\pi  a_T} - a\eta = \frac{ x \sqrt{1-\eta^2 x^2}}{x^2-1} ,  
  \end{aligned}
\end{equation}
which corresponds to a finite gap solution with branch-cuts on ${\cal C}_{{\hat 1},{\hat4} } \equiv [-\eta, \eta]$ and ${\cal C}_{{\hat 2},{\hat3} }\equiv(-\infty,-\eta^{-1}]\cup [ \eta^{-1},\infty)$ in $\mathbb{C}$.
Matching \eqref{eq:ads-qm-bmn} with the (off-shell) asymptotics \eqref{eq:QM-offshell} is consistent with the values of the conserved quantities $E=\tilde{Q}_T, \tilde{Q}_\Theta$ and $\mathbf{Q}$ given in the previous section. 
The quasimomenta of the sphere, which we denote by  $ {p}_{\tilde{1}-\tilde{4}} (x) $,   are unaffected by the deformation and have no cuts \cite{Gromov:2007aq}
\begin{equation} \label{eq:sphere-qm-bmn}
  {p}_{\tilde 1} (x) ={p}_{\tilde 2} (x)=-{p}_{\tilde 3} (x)=-{p}_{\tilde 4} (x)= \frac{2\pi a_\phi x}{x^2-1} \ .
\end{equation}

Semi-classical quantum fluctuations on top of the BMN-like solution  can be  obtained at the level of the (twisted) Spectral Curve by introducing microscopic cuts between the Riemann sheets
\eqref{eq:ads-qm-bmn}--\eqref{eq:sphere-qm-bmn} in all possible ways \cite{Gromov:2007aq,Gromov:2008ec,Borsato:2022drc}.  Heuristically, since a branch cut can be viewed as a ``condensation'' of poles, the microscopic cut or excitation can be treated as a single pole singularity. 
The excitation is bosonic when this cut connects two $AdS$ sheets ${\hat i}-{\hat j}$ 
or two sphere sheets ${\tilde i}-{\tilde j}$, while it is fermionic when it
connects an $AdS$ sheet ${\hat i}$ with a sphere sheet ${\tilde j}$, see e.g.~\cite{Beisert:2005bm}.
The  backreaction from such excitations results in shifts of the classical (background) quasimomenta as $p_i \rightarrow p_i + \delta p_i$,
$i\in \{{\hat i},{\tilde i}\}$, which must satisfy a number of analytic properties coming from the BMN-like CSC as well as from  $\mathfrak{psu}(2,2|4)$. These properties are so restrictive that they fully determine $\delta p_i$ through  a simple linear problem which we will now summarise.  

First, the corrections must not alter the gluing conditions of the classical macroscopic cuts, i.e.
\begin{equation} \label{eq:gluing-corr}
  \delta p_i (x + i \epsilon) - \delta p_j (x-i\epsilon) = 0 , \quad x\in {\cal C}_{i,j} \ ,
\end{equation}
for infinitesimal $\epsilon$. Similarly, on the location $x_n^{ij}\in\mathbb{C}$ of the new microscopic cuts  between the sheets $(i,j)$ and with mode number $n$ we have, to leading order,
\begin{equation} \label{eq:polepositions}
  p_i (x^{ij}_n) - p_j (x^{ij}_n) = 2\pi n , \qquad n\in\mathbb{Z} \ .
\end{equation}
This condition fixes the positions $x_n^{ij}$. The number $N_n^{ij}$ of such excitations  is furthermore constrained by the level-matching condition 
$\sum_n n \sum_{\text{all}~ij} N^{ij}_n = 0$.
Due to the $\mathbb{Z}_4$-grading of  $\mathfrak{psu}(2,2|4)$, the quasimomenta should in addition satisfy ``inversion symmetry'' 
\begin{equation}
  \delta p_{\hat i} (x) = -\delta p_{\hat{i}'} (x^{-1}) ,~~~\delta p_{\tilde i} (x) = -\delta p_{\tilde{i}'} (x^{-1}) , 
\end{equation}
with $i=(1,2,3,4)$ and $i'=(2,1,4,3)$,  and a ``synchronisation'' around the poles $x=\pm 1$ of the Lax connection  
\begin{equation} \label{eq:synchronisation}
  \underset{x=\pm 1}{\mathrm{Res}} \delta p_{\hat i} = \underset{x=\pm 1}{\mathrm{Res}} \delta p_{\hat{i}'} = \underset{x=\pm 1}{\mathrm{Res}} \delta p_{\tilde i}  = \underset{x=\pm 1}{\mathrm{Res}} \delta p_{\tilde{i}'} \ ,
\end{equation}
which stems, in addition, from the supertracelessness of the Lax and the Virasoro constraints. {For bosonic excitations there is a further simplification 
 ${\mathrm{Res}_{x=\pm 1}} \delta p_{\hat 1} = - {\mathrm{Res}_{x=\pm 1}} \delta p_{\hat{3}}$, and related implications from \eqref{eq:synchronisation},
due to the tracelessness of $\mathfrak{su}(2,2)$ and $\mathfrak{su}(4)$ separately.}
At last,  because of  \eqref{eq:QM-offshell}, we demand that the next-to-leading order asymptotics of $\delta p_i(x)$ around zeroes of the (gauge-transformed) Lax connection, here $x=\infty$, takes the form 
\begin{align}
  \begin{pmatrix}
    \delta p_{\hat 1} \\
    \delta p_{\hat 2} \\
    \delta p_{\hat 3} \\
    \delta p_{\hat 4} 
  \end{pmatrix}
  \sim{}& {\scalebox{1.2}{$\frac{\delta\mathbf{Q}}{2}$}} \begin{pmatrix}
    a \\ 
    i-a \\
    -i - a \\
    a
  \end{pmatrix} 
  + {\scalebox{1.2}{$\frac{4\pi}{x {\sqrt{\lambda}}}$}}
  \begin{pmatrix}
   \frac{\delta \Delta}{2} +  \sum_{\scriptscriptstyle i=\hat{3},\hat{4},\tilde{3},\tilde 4} N_{\hat{1}i} \\
   \sum_{\scriptscriptstyle i=\hat{3},\hat{4},\tilde{3},\tilde 4} N_{\hat{2}i} \\
   -\sum_{\scriptscriptstyle i=\hat{1},\hat{2},\tilde{1},\tilde{2}} N_{i\hat{3}} \\
    -\frac{\delta \Delta}{2} - \sum_{\scriptscriptstyle i=\hat{1},\hat{2},\tilde{1},\tilde{2}} N_{i\hat{4}} 
  \end{pmatrix},  \nonumber \\
  \begin{pmatrix}
    \delta p_{\tilde 1} \\
    \delta p_{\tilde 2} \\
    \delta p_{\tilde 3} \\
    \delta p_{\tilde 4} 
  \end{pmatrix}  
  \sim{}& {\scalebox{1.2}{$\frac{4\pi}{x {\sqrt{\lambda}}}$}}
  \begin{pmatrix}
    - \sum_{\scriptscriptstyle i=\hat{3},\hat{4},\tilde{3},\tilde 4} N_{\tilde{1}i}\\
   - \sum_{\scriptscriptstyle i=\hat{3},\hat{4},\tilde{3},\tilde 4} N_{\tilde{2}i} \\
   \sum_{\scriptscriptstyle i=\hat{1},\hat{2},\tilde{1},\tilde{2}} N_{i\tilde{3}}\\
   \sum_{\scriptscriptstyle i=\hat{1},\hat{2},\tilde{1},\tilde{2}} N_{i\tilde{4}}
  \end{pmatrix} \ , \label{eq:dQM-offshell-spher} 
\end{align}
where $\delta\Delta$ is the anomalous correction to the energy $E$, $\delta\mathbf{Q}$ the correction to the twist charge $\mathbf{Q}$, and $N_{ij}=\sum_n N_n^{ij}$ is the total number of excitations connecting sheets $(i,j)$ \footnote{We assume that only the twist charge and the energy can receive  corrections, while the spin charge of angular momentum $\tilde Q_\Theta$ does not, as it sits in a representation of a compact group. }. If we denote each contribution to $\delta\Delta$ from the excitation $N_n^{ij}$ as $\Omega^{ij} (x_n)$, we can write%
\begin{equation}
  \delta \Delta = \sum_n \Omega^{ij} (x_n ) N_n^{ij} \ .
\end{equation}%
In analogy with the harmonic oscillator, we will call $\Omega^{ij} (x_n)$ the  \textit{frequencies}. For more details on the origin of the properties \eqref{eq:gluing-corr}--\eqref{eq:dQM-offshell-spher} we refer to \cite{Gromov:2007aq,Gromov:2008ec,Beisert:2005bm} in the periodic case and \cite{Borsato:2022drc} in the twisted case. Note that the only difference between the periodic and twisted models lies in the asymptotics \eqref{eq:QM-offshell} and \eqref{eq:dQM-offshell-spher} and, consequently, the identification of local charges in the spectral curve. 

By combining the constraints from \eqref{eq:gluing-corr}--\eqref{eq:dQM-offshell-spher}, we can now calculate all the frequencies $\Omega^{ij} (x_n)$ as well as $\delta \mathbf{Q}$ for our model. A significant advantage of employing the s-CSC over standard semiclassical quantisation methods based on effective actions is that, for a large class of classical solutions, the full spectrum of frequencies can be completely determined from a single sphere and a single $AdS$ frequency (the ``frequency basis''). This was explicitly proven in \cite{Gromov:2008ec} for solutions with pairwise symmetric quasimomenta, i.e.~$p_{\hat{1},\hat{2},\tilde{1},\tilde{2}} = - p_{\hat{4},\hat{3},\tilde{4},\tilde{3}}$,  using inversion symmetry as well as the composition of off-shell frequencies which share poles of opposite residues. 
In our case, the (constant) $a$-dependent terms in the $AdS$ quasimomenta \eqref{eq:ads-qm-bmn} spoil this assumption; however, they only shift the reference values on the sheets and since the frequencies are derived from the $x$-dependent terms, which are pairwise symmetric, the proof in \cite{Gromov:2008ec} readily goes through. 
Consequently, using table (B.1) therein, the spectrum of frequencies can be entirely determined through, e.g., $\Omega^{\tilde 2 \tilde 3}$ and $\Omega^{\hat 2 \hat 3}$ only.

To compute the sphere frequency $\Omega^{\tilde 2 \tilde 3}$, we need to  turn on only the 
bosonic excitation $N^{\tilde 2 \tilde 3}_n$. The pole structure, incl.~residues, at $x=\pm1, x_n^{\tilde 2 \tilde 3}$ 
for e.g.~$\delta p_{\tilde 2}(x)$ is then easily obtained using inversion 
symmetry and comparison with the asymptotics \eqref{eq:dQM-offshell-spher}, similar as 
done e.g.~in \cite{Gromov:2007aq,Borsato:2022ubq}. The backreaction of this excitation on the $AdS$ sheets is expected to slightly shift the branch points connecting  $\hat{1}-\hat{4}$ and $\hat{2}-\hat{3}$,   justifying the  ansatz 
\begin{equation} \label{eq:ads14ansatz}
  \delta p_{\hat 1}(x) = f(x) + \frac{g(x)}{K(\tfrac{1}{x})} , ~~~~ \delta p_{\hat 4}(x) = f(x) - \frac{g(x)}{K(\tfrac{1}{x})}, 
\end{equation}
with  $K(x) = \sqrt{1-\eta^2 x^2}$ and $f(x), g(x)$ arbitrary functions. Using the synchronisation of the  poles at $x=\pm 1$ for this bosonic excitation and  the asymptotics \eqref{eq:dQM-offshell-spher},  we find $f(x) = a \delta \mathbf{Q}/2$ using Liouville's theorem, and 
\begin{equation}
  \delta\Delta = \sum_n \Omega^{\tilde 2 \tilde 3} (x_n^{\tilde 2 \tilde 3})  N_n^{\tilde 2 \tilde 3} , ~~~~ \Omega^{\tilde 2 \tilde 3} (x_n^{\tilde 2 \tilde 3}) = \frac{2 K(1)}{(x_n^{\tilde 2 \tilde 3})^2-1} .
\end{equation}%
~~~For any  AdS excitation, e.g.~$N_n^{\hat{2}\hat{3}}$, there will be no correction to the quasimomenta of the sphere, $\delta p_{\tilde i} (x) = 0$, which can be verified from \eqref{eq:dQM-offshell-spher} and the synchronisation of the poles at $x=\pm 1$ for bosonic excitations. We can furthermore assume the same ansatz \eqref{eq:ads14ansatz} for $ \delta p_{\hat 1}(x)$ and $ \delta p_{\hat 4}(x)$. Liouville's theorem again implies that $f(x) = a \delta \mathbf{Q}/2$  while using inversion symmetry we can write%
\begin{equation}%
  g(x) =  -\sum_n \frac{K(x_n^{\hat{2}\hat{3}}) \alpha ((x_n^{\hat{2}\hat{3}})^{-1})}{x-(x_n^{\hat{2}\hat{3}})^{-1}} N_n^{\hat{2}\hat{3}}  + ~\mathrm{reg.} , 
\end{equation}%
where $\alpha(x) = \frac{4\pi}{{\sqrt{\lambda}}} \frac{x^2}{x^2-1}$.
Matching with  \eqref{eq:dQM-offshell-spher} we  find%
\begin{equation}
  \delta\Delta = \sum_n \Omega^{\hat{2}\hat{3}} (x_n^{\hat{2}\hat{3}})  N_n^{\hat{2}\hat{3}} , ~~~\Omega^{\hat{2}\hat{3}} (x_n^{\hat{2}\hat{3}}) =  \frac{2K(x_n^{\hat{2}\hat{3}})}{(x_n^{\hat{2}\hat{3}})^2-1}
\end{equation}%
~~~As argued, using inversion symmetry and composition of the poles, the other frequencies can be easily extracted from (B.1) of \cite{Gromov:2008ec}. Note that their  form as functions of the pole positions $x_n^{ij}$ is independent from the parameter $a$ and thus coincides with (6.18)--(6.23) of \cite{Borsato:2022drc}. The expressions for $x_n^{ij}$ themselves, however, will receive $a$-contributions. After solving \eqref{eq:polepositions} for every excitation and inserting the obtained solutions for $x_n^{ij}$ in $\Omega^{ij}(x_n^{ij})$ we find%
{%
\small
\begin{align}
  \Omega^{\tilde{1}\tilde{3}} &= \Omega^{\tilde{1}\tilde{4}} = \Omega^{\tilde{2}\tilde{3}}= \Omega^{\tilde{2}\tilde{4}} 
= -\sqrt{1 - \eta^{2}} + \sqrt{1 - \eta^{2} + n^2 a_T^{-2} } , \nonumber 
  &\label{eq:final_frequencies_first}\   
  \\
  \Omega^{\hat{1}\hat{4}} &= -2 + \sqrt{2 +  n^2 a_T^{-2}  + 2\sqrt{1 + (1 - \eta^{2}) n^2 a_T^{-2}} } \nonumber 
  ,\\
  \Omega^{\hat{2}\hat{3}} &= \sqrt{2 + n^2 a_T^{-2}  - 2\sqrt{1 + (1 - \eta^{2})n^2 a_T^{-2}   }}\ , \nonumber  \\
  \Omega^{\hat{1}\hat{3}} &= -1 + \sqrt{1 + \eta^{2} + {\left(n a_T^{-1} - 2a \eta\right)^{2}}}\ , \nonumber  \\
  \Omega^{\hat{2}\hat{4}} &= -1 + \sqrt{1 + \eta^{2} + {\left(n a_T^{-1} + 2a\eta\right)^{2}}}
  \\
  \Omega^{\hat{1}\tilde{3}} &= \Omega^{\hat{1}\tilde{4}} = -1 + \sqrt{1 + {\left(n a_T^{-1} - a\eta\right)^{2}}}\ , \nonumber \\
  \Omega^{\tilde{1}\hat{4}} &= \Omega^{\tilde{2}\hat{4}} = -1 + \sqrt{1 + {\left(n a_T^{-1} + a\eta\right)^{2}}}\ ,\nonumber \\
  \Omega^{\hat{2}\tilde{3}} &= \Omega^{\hat{2}\tilde{4}}= -\sqrt{1 - \eta^{2}} + \sqrt{1 + {\left(n a_T^{-1} + a\eta\right)^{2}}}\ ,\nonumber \\
  \Omega^{\tilde{1}\hat{3}} &= \Omega^{\tilde{2}\hat{3}} = -\sqrt{1 - \eta^{2}} + \sqrt{1 + {\left(n a_T^{-1} - a\eta\right)^{2}}}\nonumber 
  \ .
\end{align}%
}%
In the undeformed limit, all of the above energy frequencies degenerate to the   single BMN frequency  \cite{Berenstein:2002jq} (then, on the Virasoro constraint, $a_T=a_\phi)$. On $a=0$ they degenerate to six independent contributions, incl.~two sets of 4-fold fermionic frequencies, that coincide with \cite{Borsato:2022drc}. 
Interestingly, we see that   turning on $a$   breaks the degeneracy. This is expected given that we  break  all  the 12 supersymmetries of the $a=0$ model 
\cite{Borsato:2022ubq}.
In the non-diagonal TsT limit,   this breaking  lifts back to five independent frequencies, incl.~two sets of 4-fold fermionic frequencies distinct from the $a=0$ case,  which can now be attributed to the restoration of the $\mathfrak{so}(6)$ symmetries. 

As a non-trivial consistency check of both the twisting and regularisation procedure,  we  verified the expressions of $\Omega^{\hat{i}\hat{j}}(x_n)$ from the periodic and deformed viewpoint through an independent calculation of the effective action of small fluctuations around the  solution \eqref{eq:bmn-like-sol}.

Let us now consider $\delta \mathbf{Q}$. First, note that we can generically write  $\delta p_{\hat 1} (x) = A + h(x)$ with $A\equiv \delta p_{\hat 1} (\infty)$ a constant and $h(x)$ a function capturing all the ${\cal O}(x^{-1})$ corrections around $x=\infty$, i.e.~$h(\infty)=0$. From the previous discussion on the $\tilde{2}\tilde{3}$ and $\hat{2}\hat{3}$ excitations, we thus have $A=a\delta \mathbf{Q}/2$ and $h(x) = g(x)/K(x^{-1})$. Therefore, in both cases,  $h(0)=0$. By inversion symmetry, this implies $\delta p_{\hat{2}}(x) =  -A + {\cal O}(x^{-1})$, which  is consistent with \eqref{eq:dQM-offshell-spher} only when $\delta \mathbf{Q}=0$. This can be readily generalised to all the other excitations, as we can expect from their composition rules. We thus obtain no anomalous correction to the twist charge, which matches with the $a=0$ Jordanian string \cite{Borsato:2022drc} and the $\beta$-deformation \cite{Beisert:2005if}. 

At last, we can   compute the one-loop correction to the vacuum energy of our classical string  \cite{Frolov:2002av,Gromov:2007aq}
\begin{equation}
  E \approx Q_T + E_{\text{\tiny 1-loop}} = Q_T + \frac{1}{2} \sum_{n\in\mathbb{Z}} \sum_{ij}(-)^{F_{ij}} \Omega_n^{ij} ,
\end{equation}
with $F_{ij}=0$ for bosonic and $F_{ij}=1$ for fermionic excitations. After approximating the sums as integrals by assuming (without loss of generality) $a_T\gg 1$, and using similar integration tricks as in \cite{Borsato:2022drc}, we find that the $a$-dependency drops out in the result after a number of non-trivial cancellations. Hence, we obtain 
\begin{equation}
   \begin{aligned}
E_{\text{\tiny 1-loop}} ={}& a_T \left(  \eta (1-\eta) - (3-\eta^2)\log\sqrt{1-\eta^2} \right. \\ & ~~~\left.  -(1+\eta^2)\log\sqrt{\frac{(1+\eta)(1+\eta^2)}{1-\eta}} \right) \ , 
 \end{aligned}
\end{equation}
as for the $a=0$ Jordanian string \cite{Borsato:2022drc}. Interestingly, this means that in the non-diagonal TsT limit $E_{\text{\tiny 1-loop}}$ vanishes despite the broken supersymmetry.

\vspace{10pt}

\begin{acknowledgments}
\textit{\textbf{Acknowledgements:}}
We thank  Niklas Beisert, Martí Berenguer Mimó, Sylvain Lacroix, Stijn Van Tongeren, and especially Riccardo Borsato, for useful discussions and comments on the draft. This work is partly based on the Master thesis of one of the authors (NK) prepared at ETH Z\"urich. SD is  supported by the Swiss National Science Foundation through the NCCR SwissMAP. NK is supported in part by the FWO Vlaanderen through the project G006119N, as well as by the Vrije Universiteit Brussel through the Strategic Research Program “High-Energy Physics”.
\end{acknowledgments}

\newpage

\bibliographystyle{apsrev4-1}
\bibliography{ads5-lcgf}

\end{document}